\pdfoutput=1 
\documentclass[%
reprint,
superscriptaddress,
 amsmath,amssymb,
 aps,
]{revtex4-1}

\usepackage{graphicx}
\usepackage{physics}
\usepackage{braket} 
\usepackage{amsfonts}
\usepackage{dsfont}
\usepackage{xcolor}
\usepackage{enumitem}
\usepackage{soul}
\graphicspath{{"../figures/"}}
\usepackage{textcomp}
\usepackage{placeins}
\usepackage[normalem]{ulem}

\usepackage{siunitx}
\DeclareSIUnit\dBm{dBm}

\usepackage{hyperref}

\begin{document}

\title{Q-BiC: A biocompatible integrated chip for \textit{in vitro} and \textit{in vivo} spin-based quantum sensing}

\author{Louise Shanahan}
\altaffiliation[]{These authors contributed equally}
\author{Sophia Belser}
\altaffiliation[]{These authors contributed equally}
\author{Jack W. Hart}
\altaffiliation[]{These authors contributed equally}
\author{Qiushi Gu}
\author{Julien R. E. Roth}
\author{Annika Mechnich}
\author{Michael H\"ogen}
\author{Soham Pal}
\affiliation{Cavendish Laboratory, University of Cambridge, JJ Thomson Avenue, Cambridge, CB3 0HE, United Kingdom}
\author{David Jordan}
\author{Eric A. Miska}
\affiliation{Department of Biochemistry, University of Cambridge, Cambridge, CB2 1GA, United Kingdom}
\author{Mete Atat\"ure}
\altaffiliation[Correspondence to: ]{ma424@cam.ac.uk, hsk35@cam.ac.uk}
\author{Helena S. Knowles}
\altaffiliation[Correspondence to: ]{ma424@cam.ac.uk, hsk35@cam.ac.uk}
\affiliation{Cavendish Laboratory, University of Cambridge, JJ Thomson Avenue, Cambridge, CB3 0HE, United Kingdom}

\begin{abstract}
Optically addressable spin-based quantum sensors enable nanoscale measurements of temperature, magnetic field, pH, and other physical properties of a system. 
Advancing the sensors beyond proof-of-principle demonstrations in living cells and multicellular organisms towards reliable, damage-free quantum sensing poses three distinct technical challenges. First, spin-based quantum sensing requires optical accessibility and microwave delivery. Second, any microelectronics must be biocompatible and designed for imaging living specimens. Third, efficient microwave delivery and temperature control are essential to reduce unwanted heating and to maintain an optimal biological environment. Here, we present the Quantum Biosensing Chip (Q-BiC), which facilitates microfluidic-compatible microwave delivery and includes on-chip temperature control. We demonstrate the use of Q-BiC in conjunction with nanodiamonds containing nitrogen vacancy centers to perform optically detected magnetic resonance in living systems. We quantify the biocompatibility of microwave excitation required for optically detected magnetic resonance both \textit{in vitro} in HeLa cells and \textit{in vivo} in the nematode \textit{Caenorhabditis elegans} for temperature measurements and determine the microwave-exposure range allowed before detrimental effects are observed. In addition, we show that nanoscale quantum thermometry can be performed in immobilised but non-anaesthetised adult nematodes with minimal stress. These results enable the use of spin-based quantum sensors without damaging the biological system under study, facilitating the investigation of the local thermodynamic and viscoelastic properties of intracellular processes.
\end{abstract}

\maketitle
\section{Introduction}

Quantum sensing based on optically addressable spins offers a promising route to probe a variety of properties of biological systems, including temperature \cite{Neumann2013}, magnetic field \cite{Horowitz2012} and pH \cite{Rendler2017}, with high sensitivity and nanoscale-spatial resolution. Several room temperature quantum sensors have been explored in recent years including lattice defects in diamond \cite{Gruber1997, Belser2023}, silicon carbide \cite{Kraus2014} and hexagonal boron nitride \cite{Gottscholl2021}. Nitrogen-vacancy (NV) centers, which are comprised of a substitutional nitrogen atom with a neighbouring vacant lattice site, have emerged as a leading candidate for quantum sensing. Nanosized diamond probes containing NVs offer a high spatial resolution and have been shown to be non-cytotoxic \cite{Zhu2012}, amenable to surface functionalisation \cite{Krueger2012,Shenderova2015} and can be introduced into a variety of living systems \cite{hebisch2021nanostraw, vaijayanthimala2009biocompatibility, Choi2020}. The NV-associated electronic spin can be probed using optical and microwave excitation, where changes in the spin energy levels indicate a change in external parameters. This provides a unique opportunity to improve our understanding of biological processes such as cell division \cite{Choi2020} and intracellular transport \cite{Gu2023}. It further allows us to address unanswered questions relating to biological mechanisms at the nanoscale \cite{McCoey2020} and quantum biology \cite{Hore2016, Buchachenko2019}. 
Spin-based quantum probes require the delivery of the microwave frequency magnetic field to the region of interest. While the biocompatibility of nanodiamonds has been studied \cite{Wodhams2018,Zhu2012, Mohan2010}, the biocompatibility of quantum sensing itself and in particular the effect of exposure to microwave excitation remains unknown. Water strongly absorbs energy in the 2-4 GHz microwave frequency range relevant for NVs, causing dielectric heating \cite{Lunkenheimer2017}. Microwave irradiation can change the local temperature profile and can have a detrimental effect on the dynamics of live specimens \cite{banik2003bioeffects}. Thus, controlling and understanding temperature variations is crucial to providing biocompatible environments.

The challenge of microwave delivery to biological specimens is not limited to the biocompatibility of microwave exposure, but also includes the precise spatial delivery of microwave fields with minimal assembly time. Firstly, spin-based quantum biosensing modalities rely on the reproducible delivery of a spatially uniform field profile to the specimen. The delivery of microwave is well established in solid-state systems and has so far been achieved with a wide variety of design implementations, including simple wires \cite{maze2008nanoscale}, external stereoscopic antennae \cite{chen2018large} and co-planar structures written by optical lithography \cite{horsley2018microwave, Oshimi2022}. However, combining microwave delivery with living specimens in a biologically stress-free, natural environment remains challenging. Biological specimens can vary dramatically in size, from a few microns for bacteria or yeast cells, to a few millimeters in small animals such as \textit{Caenorhabditis elegans}, which must be accommodated for in the sensing setup. The biological specimens of interest must be reproducibly located in an optically accessible imaging region in close proximity to the microwave source. For cells, this can be achieved with adhesive coatings. However, more complex biological specimens, such as \textit{C. elegans}, require sophisticated immobilisation for alignment along the microwave line. Preferred culture environments also differ significantly between species. For example, \textit{C. elegans} require a liquid immersion film \cite{Stiernagle2006}, whereas cell cultures need to be submerged fully in a liquid culture medium \cite{KumarMicrobes, Spencer}. These media are often electrolytic, leading to incompatibility with electronics and causing additional challenges, like surface deterioration. Finally, minimal assembly time is essential, as the timescales involved in the biological experiments of interest are significantly shorter than those of abiotic quantum experiments, where there is usually no timescale of concern associated with degradation. 

\begin{figure}
\includegraphics[width=0.48\textwidth]{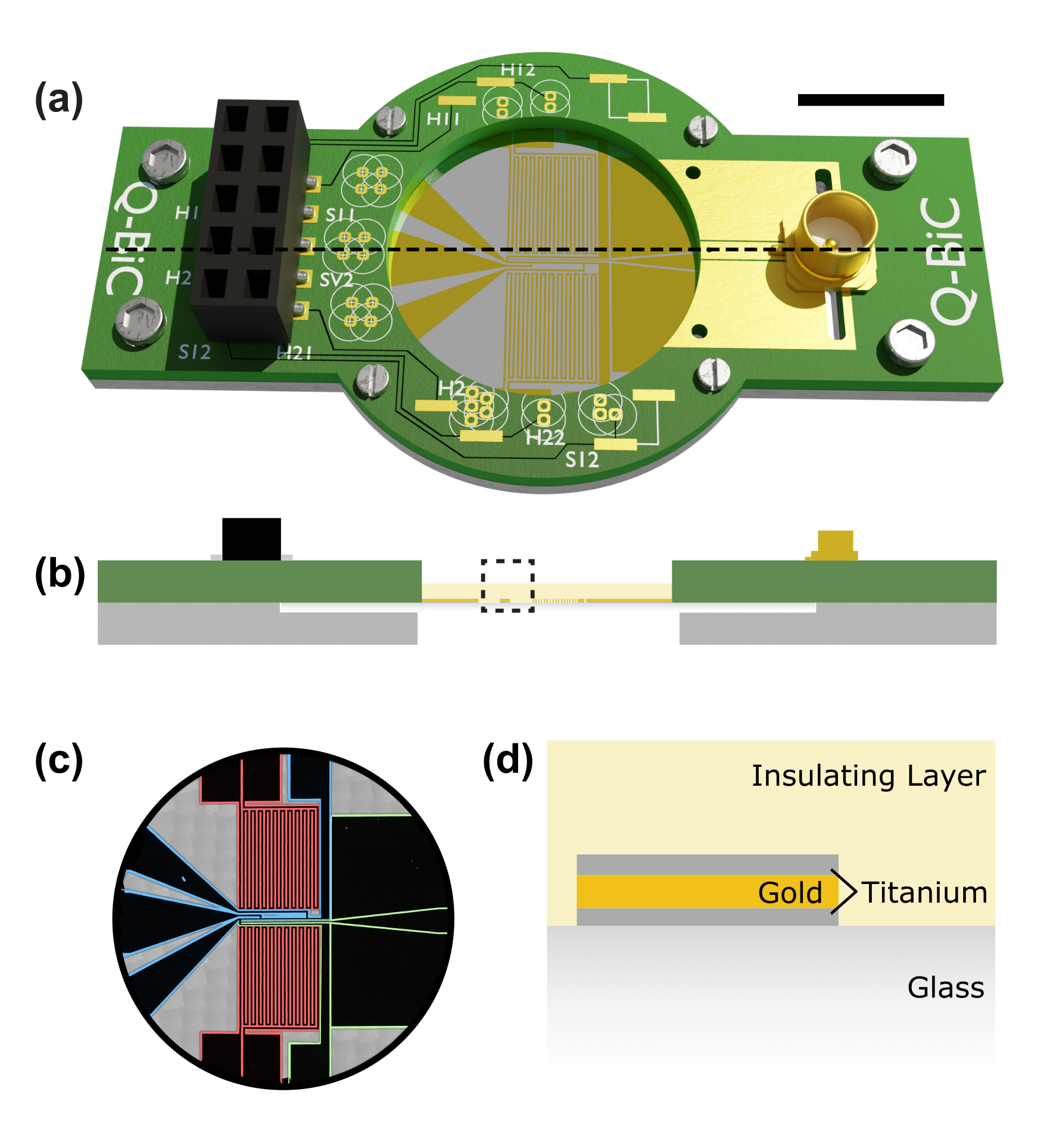}
\caption{\textbf{(a)} Top down view of chip showing the black connector for heating and temperature sensing and the SMP connector for Microwave delivery. Scalebar: 1 cm. \textbf{(b)} Cross section of chip (not to scale) along the dotted line in (a) showing glass sandwiched between PCB (green) and aluminium mount (grey). \textbf{(c)} Confocal image of the glass chip showing the microwave lines (green), heaters (red) and RTD (blue). \textbf{(d)} Cross section of glass slide presented in (b). Layers of titanium, gold and titanium are evaporated onto the glass slide which is coated with a layer of PDMS or Parylene C for insulation.
\label{Fig1}}
\end{figure}

In this work, we present Q-BiC to address the challenges of biological quantum sensing. Q-BiC is a compact microelectronic chip, which integrates global temperature sensing and temperature modulation, microwave delivery, and microfluidics-compatibility as shown in Fig. \ref{Fig1} \textbf{(a)}. To facilitate recording the optical readouts of quantum sensors, as well as those of fluorescence-based microscopy techniques common in bio-imaging, Q-BiC has been designed with a clear imaging area $150\,\mathrm{\mu m} \times 5\,\mathrm{mm}$. The chip is fabricated in a simple photolithography process and is easily assembled. It is also compatible with sterilisation processes allowing for re-use and use with potentially harmful biological material. We quantify the uniformity of the microwave delivery by measuring the field strength in the vicinity of the antenna and showing that it is in good agreement with predictions from simulations. We demonstrate Q-BiC's ability to control the specimen temperature and highlight the importance of being able to measure the specimen's global temperature while looking for nanoscale temperature responses. We show how liquid cell cultures can be integrated on the chip and study the heating effects caused by microwave excitation. We demonstrate quantum sensing using NVs in HeLa cells and the biocompatibility thereof, identifying the threshold for microwave excitation power that allows for stress-free operation. Further, we present a method to immobilise live \textit{C. elegans} to a region of interest, which enables us to demonstrate optically detected magnetic resonance (ODMR) measurements directly in a non-anaesthetised adult animal. Finally, taking advantage of a genetically encoded fluorescent stress reporter in \textit{C. elegans}, we demonstrate the biocompatibility of quantum sensing in live nematodes.

\section{Results}

\begin{figure*}[htbp]
\includegraphics[width=1\textwidth]{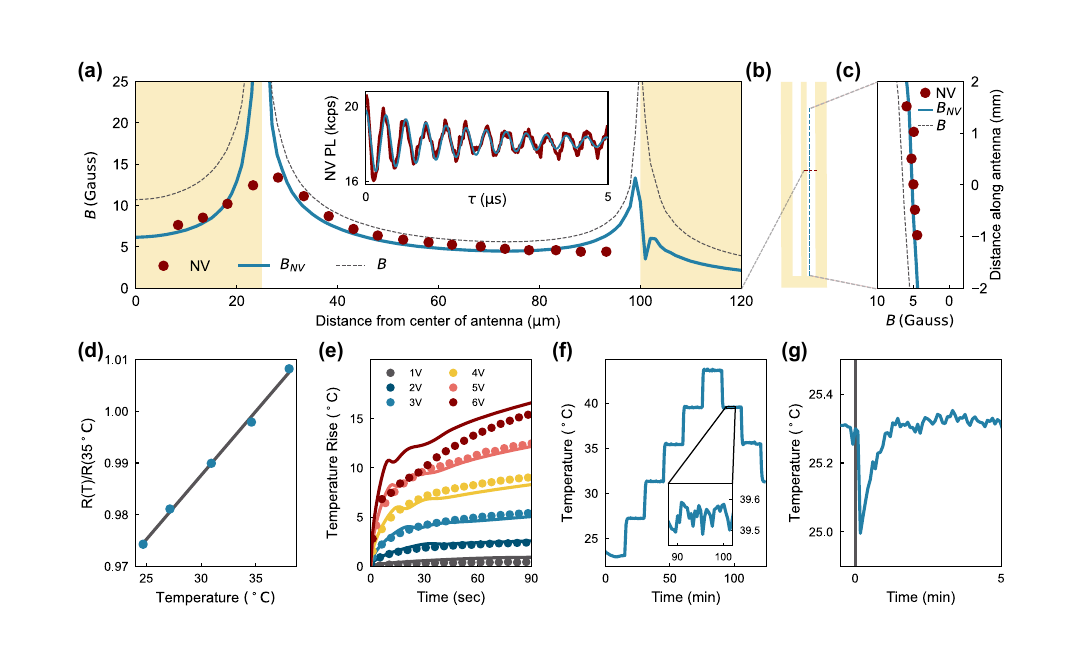}
\centering
\caption{\textbf{(a)} Simulation of the magnetic field (grey dashed curve) along the transverse cross sections of the measurement region. The experimental data measured by a diamond tip (solid red circles) is consistent with the projection of the magnetic field simulation orthogonal to the NV axis (blue curve). \textbf{Inset:} Rabi oscillations between $m_s = 0$ and $m_s = \pm1$ states. \textbf{(b)} Schematic of the measurement region of the chip showing the transverse cross section (red dashed) and longitudinal cross section (blue dashed) relative to the gold antenna (yellow). \textbf{(c)} Simulation of the magnetic field (grey dashed curve) along the blue dashed line in (b). The experimental data measured by a diamond tip and projection of the magnetic field simulation orthogonal to the NV axis can be seen as solid red circles and the blue curve, respectively. \textbf{(d)} Calibration of the on-chip thermometer. The resistance of the RTD, \textit{R}, at a temperature, \textit{T}, relative to the resistance at a known reference temperature, $R(T_{\mathrm{ref}})$, is proportional to the temperature, \textit{T}, where the proportionality constant, $\eta$, is related to the temperature coefficient of gold. \textbf{(e)} The temperature increase reported by the on-chip RTD when different voltages are applied across the contacts of the on-chip heaters. At low voltages the experimental data (solid circles) is seen to agree with simulations (solid curves). \textbf{(f)} Demonstration of controlled temperature stepping. \textbf{Inset:} The PID is shown to stabilise the temperature to within $\sigma<30\,\mathrm{mK}$. \textbf{(g)} Characterisation of external or unwanted sources of heat. A transient temperature decrease is observed when $50\,\mathrm{\mu L}$ of water stored at room temperature is added to a sample containing $400\,\mathrm{\mu L}$ of water. The errors on the experimental data in \textbf{(a)}, \textbf{(c)} and \textbf{(d)} are smaller than the marker size.}
\label{Fig2}
\end{figure*}

\subsection{Experimental characterisation and simulations of the magnetic field distribution}
The reliable delivery of microwave excitation is a challenge for spin-based quantum sensing in biological systems. Ideally, a sensing chip would provide a uniform field distribution over a large imaging region and require minimal assembly time post addition of the biological specimen. We have integrated a co-planar waveguide (CPW) into Q-BiC to address this challenge, allowing for reproducible microwave delivery and rapid operation after specimen addition.

The microwave field is delivered to Q-BiC via a $50\,\Omega$-matched CPW on FR4 fiberglass PCB substrate. This is connected to the microwave source using a subminiature push-on (SMP) cable with all soldering connections carried out prior to the introduction of the biological specimen, as seen in Fig. \ref{Fig1} \textbf{(b)}. The CPW continues on the glass substrate and tapers at the central sample region as shown in Fig. \ref{Fig1} \textbf{(c)} (green). The sample region consists of a $5\,\mathrm{mm}$-long, $50\,\mathrm{\mu m}$-wide microwave line with $70\,\mathrm{\mu m}$ of clear imaging region on either side. The imaging region is insulated from the CPW using a $5\,\mathrm{\mu m}$ layer of Polydimethylsiloxane (PDMS) or a $2\,\mathrm{\mu m}$ Parylene C layer as shown in Fig. \ref{Fig1} \textbf{(d)}. Both coatings have low autofluorescence (Supplementary Figure S1). 

We quantify the field distribution in this clear imaging region using a scanning NV microscope and compare these results with numerical simulations as seen in Fig. \ref{Fig2}. The scanning NV microscope employs a single NV close to the apex of a diamond tip attached to a tuning fork which gives nanometer-scale spatial resolution. The diamond tip is cut along the [110] plane such that the NV symmetry axis is tilted $30^{\circ}$ with respect to the sample plane, perpendicular to the CPW orientation. As seen in Fig. \ref{Fig2} \textbf{(a) inset}, we measure the Rabi frequency of this NV. The Rabi frequency, $\Omega$, relates to the local magnetic field orthogonal to the NV axis, $B_{\mathrm{NV}}$, as 

\begin{equation}
    \Omega = \frac{1}{2}\mu_\mathrm{B} g \mathrm{B}_{\mathrm{NV}}/\hbar.
\end{equation}

\noindent where $\mu_B$ is the Bohr magneton, \textit{g} is the g-factor for an electron and $\hbar$ is the reduced Planck constant \cite{Du2012}. Figure \ref{Fig2} \textbf{(a)} and \textbf{(c)} (solid red circles) show the field distribution measured by the scanning NV microscope tip perpendicular to and along the length of the CPW as demonstrated by the red and blue dashed lines in Fig. \ref{Fig2} \textbf{(b)}. The relationship between the Rabi frequency and the microwave power is shown in Supplementary Figure S2. We compare these measurements with simulations which are performed using the finite element method (See Methods). These simulations give us the total magnetic field (grey dashed curve) and the projection of the magnetic field simulation orthogonal to the NV axis (blue curve) which shows good agreement with the experimental results. We note in particular that resistive losses in the metallic structure were considered in the modelling. Due to the strong confinement of the electric field, the simulation matches the measured field without considering the presence of the metal objective and auxiliary aluminium mounting.

\subsection{Temperature Sensing and Heating} 
The Q-BiC combines the delivery of microwave with temperature control over a timescale of minutes and an on-chip temperature sensor that can monitor the global specimen temperature. The temperature of the local environment should be stable over long periods of time, to counteract environmental drifts. Additionally, the ability to tune the temperature is essential for calibration purposes. 

The temperature of the sensing chip is set using two resistive heaters which are regulated using feedback from the resistive temperature detector (RTD). The operation of the RTD is based on the linearity of electrical resistance of gold over the relevant temperature range, 20 - 50 $^{\circ}$C. As seen in Fig. \ref{Fig2} \textbf{(d)} the resistance, \textit{R}, of the RTD is linearly proportional to temperature, \textit{T}. This is described by

\begin{equation}\label{equation: RTD}
    R(T)/R(T_{\mathrm{ref}}) = \eta (T-T_{\mathrm{ref}}) + 1,
\end{equation}

\noindent where $R(T_{\mathrm{ref}})$, is the resistance at a known reference temperature and $\eta$ is an experimentally determined constant related to the temperature coefficient of gold. $\eta$ is characterised for five separate chips using a temperature controlled incubator box and is shown to have less than $5\%$ variation between chips (Supplementary Figure S3). 

The two on-chip heaters generate Joule heating, which in turn causes the sensing chip to heat up through thermal transfer. The increase in temperature, measured by the on-chip RTD when a current is passed through the heaters for different voltages, $V_0$, can be seen in Fig. \ref{Fig2} \textbf{(e)}. The temperature of fluid in the channel can be described as 

\begin{equation}
    \label{equation:heaters}
    C_v \frac{dT}{dt} = e\frac{V_0^2}{R_\mathrm{heater}} - k(T-T_0),
\end{equation}

\noindent where $C_v$ is the (constant volume) heat capacity of the fluid, \textit{e} is the duty cycle, $R_\mathrm{heater}$ is the resistance of the heater, $T_0$ is the room temperature and \textit{k} is the rate of heat loss due to thermal conduction. It should be noted that this model only takes into account heat loss due to thermal conduction. At higher voltages, convection may cause additional heat loss and our model would over estimate the temperature increase (Supplementary Figure S4). The measured temperature increase seen in Fig. \ref{Fig2} \textbf{(e)} (filled circles)  shows good agreement with the heat generation simulation (solid curve, see methods).

Equation \ref{equation:heaters} can be used to generate and tune a Proportional Integral Derivative (PID) control algorithm for optimal control of the on-chip temperature, based on the temperature measured by the on-chip RTD. Figure \ref{Fig2} \textbf{(f)} demonstrates the PIDs ability to perform controlled steps in temperature, a useful feature for calibration of the quantum sensors. Temperature stepping over a shorter timescale is demonstrated in Supplementary Figure S5. The PID control stabilises the temperature of the sensing chip to within \SI{30}{\milli\kelvin} as seen in Fig. \ref{Fig2} \textbf{(f) inset}.

One of the challenges in temperature sensing is external or unwanted sources of heat which can lead to erroneous results. Therefore all potential sources of heat need to be characterised. An example of this is the addition of liquids to the specimen. As seen in Fig. \ref{Fig2} \textbf{(g)}, the on-chip RTD measures the global temperature change after the addition of a chemical which is stored at a different temperature to the specimen. This is crucial to identify, so that global changes in temperature are not incorrectly attributed to the chemical reactions or to the nanoscale fluctuations we are attempting to characterise using the NV quantum sensor.

\begin{figure*} [htbp!]
\centering
\includegraphics[width=1\textwidth]{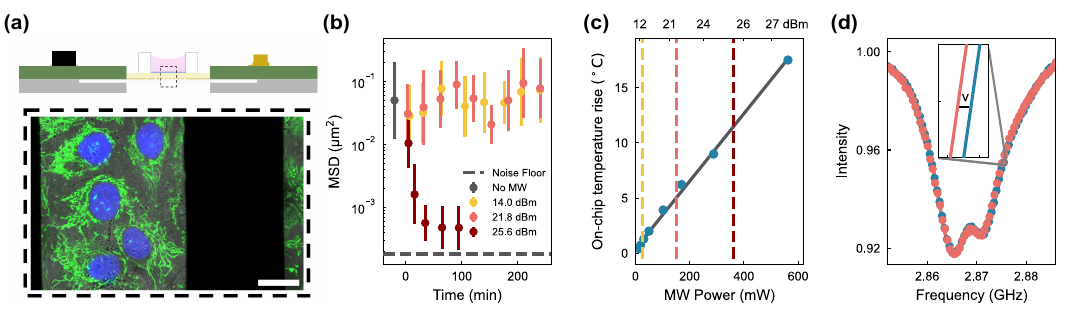}
\caption{ \textbf{(a)}  Visualisation of the Q-BiC with an attached PDMS well for cell culture. Inset: Fluorescence image of HeLa cells growing along the microwave line (green = MitoTracker Green FM, blue = NucBlue, scalebar = 20 $\mu$m).  \textbf{(b)} Cell Viability. Time course data showing the mean and standard deviation of the mean squared displacement at a time interval of 10 seconds for intracellular vesicles under continuous microwave excitation. 14.0 dBm and 21.8 dBm were shown to have no effect on particle movement over the course of 4 hours, indicating good cell health. Microwave at 25.6 dBm were shown to cause a dramatic decrease in particle movement over approximately 30 minutes resulting in cell death at 100 min. The noise floor of the tracking system is plotted as the dashed grey line. \textbf{(c)} The on-chip temperature rise observed for different microwave powers (solid blue circles) with the best linear fit (grey curve). The microwave powers used in (b) are highlighted as dashed lines. \textbf{(d)} Safe delivery of microwave enables the use of spin based quantum sensing. The intracellular temperature can be read out by measuring the shift in the optically detected magnetic resonance spectrum (inset). This ODMR spectrum is seen to shift from higher frequencies (blue curve) to lower frequencies (red curve) when the temperature is increased by $2\, ^{\circ}\mathrm{C}$.
\label{Fig3}}
\end{figure*}

\subsection{\textit{In vitro} and \textit{in vivo} calibration}
\subsubsection{Quantum sensing in HeLa cells}
The energy added to a specimen upon microwave excitation is known to cause heating, with reports of temperature increases by as much as 16 K \cite{Wang2022}. For biological specimens, which are predominantly water-based, strong microwave excitation has been associated with excessive heating and cell death \cite{Asano2017,Lunkenheimer2017,banik2003bioeffects}. To assess cell viability during continuous microwave exposure, we measured the mean squared displacement (MSD) of intracellular vesicles in HeLa cells on PDMS-coated Q-BiCs at an ambient temperature of $37 \, ^{\circ}\mathrm{C}$ (Fig. \ref{Fig4} (a)). Cells were considered healthy if the average MSD at a time interval of 10 seconds did not change significantly in response to the applied microwave, as intracellular viscosity has been shown to increase upon cell death \cite{kuimova2009imaging} (See Methods and Supplementary Figure S6). Cells were exposed to microwave for up to 4 hours to test for any long-term heating effect. As seen in Fig. \ref{Fig4} (b), microwave power (measured before connection to the antenna) up to 21.8 dBm showed a negligible effect on MSD compared to average MSD when no microwave was applied, suggesting no detrimental effect to cell health. When the microwave power was increased to 25.6 dBm, the MSD was seen to decrease dramatically over the first 20 minutes. At this microwave power, cells within $300\,\mathrm{\mu m}$ of the microwave antenna were killed, as evidenced by the retention of trypan blue solution (Supplementary Figure S7).

The heating effect caused by microwave excitation is the likely cause of cell death. The magnitude of the resulting temperature rise is dependent on the power of the microwave excitation, the conductivity of the specimen and the background temperature. Figure \ref{Fig4} \textbf{(c)} shows the temperature increase measured by the on-chip RTD as a function of applied microwave excitation power. The specimen consisted of $500\, \mathrm{\mu L}$ of water in a PDMS well, and the temperature was measured for 20 minutes after the microwave were turned on. The increase in specimen temperature occurred predominantly in the first 10 seconds after the microwave was applied and the specimen was seen to reach thermal equilibrium after approximately 5 minutes (Supplementary Figure S8). These microwave induced temperature increases can be compensated for by lowering the incubator temperature and fine-tuned using the on-chip temperature regulation.

Figure \ref{Fig4} \textbf{(d)} shows ODMR readout from an ensemble-NV nanodiamond at $24\, ^{\circ}\mathrm{C}$ and $26\, ^{\circ}\mathrm{C}$. These measurements were taken with a microwave power of $19\,\mathrm{dBm}$ which is below the power threshold we measured to be detrimental to cell health. This microwave power leads to a temperature sensitivity of 1.4 $\mathrm{K}/\sqrt{\mathrm{Hz}}$ \cite{Gu2023}, which is sufficient to observe thermogenesis in a cellular environment \cite{di2021quantitatively}. This demonstrates the ability to perform biocompatible quantum sensing \textit{in vitro} using Q-BiC.

\subsubsection{Quantum sensing in \textit{C. elegans}}
\begin{figure*}[htbp!]
\includegraphics[width=1\textwidth]{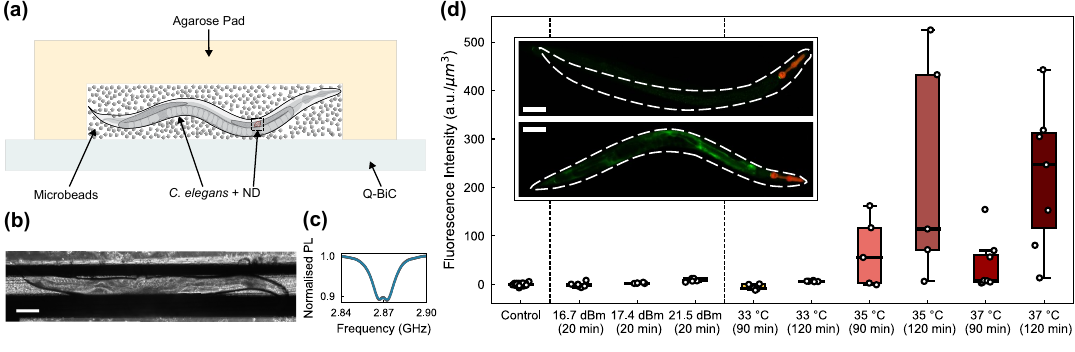}
\caption{Sensing in live \textit{C. elegans} on Q-BiC. \textbf{(a)} Schematic of the nanoparticle-mediated immobilisation of live \textit{C. elegans} adults. A nanodiamond (not to scale, highlighted with dashed box) is shown in the distal arm of the gonad, the location of micro-injection.  \textbf{(b)} Phase-contrast image of an immobilised live adult nematode on Q-BiC alongside the microwave line (left). Scale-bar = 100 $\mu$m. \textbf{(c)} Example ODMR spectrum taken in immobilised \textit{C. elegans} as shown in (a), proving the possibility of spin based quantum sensing applications in live worms without the use of anaesthetics (21 dBm). \textbf{(d)} Stress level of worms immobilised for 20 min with active microwave at different powers (16.7 dBm with an effective temperature of 25.1 °C, 17.4 dBm with an effective temperature of 25.7 °C and 21.5 dBm with an effective temperature of 31.3 °C in comparison with the stress induced by exposure to 33 °C, 35 °C and 37 °C, for 90 min and 120 min each. The fluorescent intensity was normalised with respect to the worm volume (division) and to the mean fluorescence of the control groups (subtraction). Inset: Confocal z-sum-projections of a non-stressed control worm (top) and a stressed worm (37 °C, 90 min) expressing GFP (bottom). Scale-bar = 100 $\mu$m.}
\label{Fig4}
\end{figure*}

Multimodal quantum sensing of temperature and viscosity have previously been demonstrated \textit{in vitro}, revealing local viscoelestic properties of subcellular environments and linking them to their nanoscale thermal landscape \cite{Gu2023}. Another opportunity for this technique is to understand complex biological processes in real time, in living organisms. 

\textit{C. elegans} are a soil nematode, commonly described as a ‘model biological system’ due to their rapid life cycle of three days, their defined cell lineage and the straightforward lab cultivation. Most importantly for optical sensing, \textit{C. elegans} are fully transparent. The transparency of the body, the eutelic nature of the organism as well as the constant cell positioning across all worms offer further advantages over other biological systems. One challenge for highly sensitive \textit{in vivo} quantum sensing is that \textit{C. elegans} can move at speeds on the order of mm/s \cite{Jung2016-lq}. Therefore, the nematodes need to be macroscopically immobilised to allow for a reliable read-out. This is often done by use of fixation solutions, like paraformaldehyde (4 \% in PBS), which can cause structural anomalies in metabolic proteins \cite{Kim2017}. Anaesthetics, such as the metabolic inhibitor sodium azide, are also commonly used for immobilisation \cite{Manjarrez2020, Fujiwara2020, Oshimi2022, Choi2020}. However, sodium azide is a potent inhibitor of mitochondrial respiration known to induce drastic physiological changes \cite{Bowler2006,HORECKER1948,smith1991acute}. The role of mitochondrial function is of particular interest for nanodiamond thermometry, as it has the potential to provide a resolution of the ``hot mitochondria paradox", a five order of magnitude discrepancy between theoretical predictions and measurements of temperature gradients near mitochondria \cite{Macherel2021-am,Chretien2018-dd,Baffou2014-zx}. Therefore, immobilisation methods that interfere with mitochondrial respiration can be problematic. We overcome this problem by employing a reliable nanoparticle-mediated immobilisation protocol that allows the immobilisation of live, non-anaesthetised \textit{C. elegans} alongside the microwave line of Q-BiC (Fig. 5 (a), (b), Supplementary Figure S9) \cite{Kim2013}. Using this immobilisation for \textit{C. elegans} which have been injected with nanodiamonds, we can obtain an ODMR read-out in live animals without having to interfere with metabolic activities of the organism (Fig. 5 (c)). We can use both PDMS (Supplementary Figure S10) or Parylene C coated Q-BiCs for sensing in \textit{C. elegans}, but will use Parylene C in the following as it can be cleaned more reproducibly between different worm mounts and is more durable when reusing the chips.
To ensure that the immobilisation and microwave exposure do not have detrimental effects on the well-being of the nematode, we use a fluorescent reporter (strain SX2635), which enables a visual read-out of stress in individual \textit{C. elegans} \cite{LePen2018}. The sensor is comprised of a fluorescent protein (GFP) that is under the control of a promoter (\textit{lys-3}), which leads to GFP expression in response to stress in general and in particular to viral infection \cite{Pirenne2021}. Further, a second  fluorescent protein (mCherry) is constitutively expressed by a tissue-specific promoter (myo-2) in the nematode pharynx, serving as a genotype control (Supplementary Figure S11). To understand the relationship between temperature and stress, the nematodes were exposed to elevated temperatures using an incubator. The stress experienced due to this heat shock is shown in Fig. \ref{Fig4} (d). The absolute fluorescent signal of \textit{C. elegans} was obtained through confocal imaging and was normalised with respect to the total volume of respective animals by division and with respect to the mean of the daily control group through subtraction (Supplementary Figure S10, S12, S13). We show that our \textit{lys-3p::gfp} sensor is capable of capturing stress induced by heat, as demonstrated by the fluorescent response of \textit{C. elegans} exposed to different temperatures for 90 min and 120 min.
We assess the impact of live immobilisation and of additional microwave exposure on the well-being of \textit{C. elegans} by comparing their fluorescence to that of heat stress. Figure \ref{Fig4} (d) shows that immobilisation with exposure to lower microwave powers (16.7 dBm) do not cause any stress for specimens. Increasing the microwave power to 21.5 dBm did not show any significant stress for the animal. Stepping up the microwave power to 22.3 dBm lead to a 100\% lethality (n=6), as opposed to 5\% in all other microwave conditions combined (n=21) (Supplementary Table S1)
In order to give context to the stress caused by microwave exposure, we determine an ``effective temperature" experienced by the specimens. It is comprised of the room temperature at the time of stress plus the temperature increase due to microwave exposure. 16.7 dBm was determined to correspond to an effective temperature of 25.1 °C, 17.4 dBm to 25.7 °C, 21.5 dBm to 31.3 °C and 22.3 dBm to 33.3 °C (Supplementary Figure S14). 
As a result, we can conclude that our immobilisation technique and microwave exposure required for reliable \textit{in vivo} nanodiamond sensing on Q-BiC does not have harmful effects on the health of \textit{C. elegans} that can be detected through induction of the \textit{lys-3} promoter. Qauntum sensing can safely be performed as long as we remain below the identified threshold of 21.5 dBm microwave excitation power. 

\section{Conclusion}
Our Q-BiC is a ready-to-use, biocompatible and reusable quantum sensing chip with full temperature control and ODMR capability for unicellular and non-anaesthetised multicellular biological specimens. 
It provides two key advances for quantum sensing in life sciences: Firstly, the biological specimen can remain intact during quantum sensing. Secondly, Q-BiC
allows the distinction between extrinsic (Q-BiC RTD) and intrinsic (NV sensor) changes in measured temperature. This is key the for reliable identification of metabolic activity through measured changes in subcellular temperature.

We characterised the performance of the chip by measuring the magnetic field distribution on the chip as well as the heating effects and readout from the on-chip heaters and RTD respectively. To asses the biocompatibility of quantum sensing \textit{in vitro} using Q-BiC, we quantified the resulting heating as well as the stress experienced by HeLa cells upon microwave exposure. We further demonstrate the precise immobilisation of non-anaesthetised \textit{C. elegans} on Q-BiC and show \textit{in vivo} biocompatibility of quantum sensing at different microwave powers.

The next steps in biocompatible quantum sensing include targeting different organelles in cells/nematodes and incorporating the multi-modal operation \cite{Gu2023} to study how nanoscale subcellular heterogeneities arise and how they affect cell function. These include temperature gradients near mitochondria that arise from respiration or viscosity differences that may result from the formation and ageing of phase separated granules. The extensive heterogeneity of the cytoplasm is only now being realised \cite{Garner2023-ij}, and biocompatible quantum sensing will be an invaluable tool for studying the physics that govern it and its implications for cell biology.
In addition to surface functionalisation, the implementation of additional quantum sensing protocols, including $T_1$ spin relaxation or Spin Echo Double Resonance (SEDOR), will be integral to exploring additional biological features such as ROS production \cite{Belser2023}.

\section{Acknowledgements}
We would like to thank Noah Shofer and Sophie Oldroyd for technical support, and Dr. John Jarman and Dr. Hannah Stern for useful discussions.
This work was supported by the Gordon and Betty Moore Foundation (GBMF7872), the Isaac Newton Trust (23.23(j)), Cancer Research UK RadNet Cambridge (C17918/A28870) and Cancer Research UK (11832) to E.A.M., the Pump Priming Grant from the Cambridge Centre for Physical Biology, Wellcome United Kingdom (104640, 207498, 0292096); as well as by the Royal Society through a University Research Fellowship held by H.S.K.. L.S. acknowledges the financial support from the Winton Programme for Sustainability and the Robert Gardiner Memorial Scholarship. S.B. acknowledges financial support from EPSRC (PhD Studentship EP/R513180/1). Q.G. acknowledges financial support by the China Scholarship Council, the Cambridge Commonwealth, European \& International Trust.\\

\section{Methods}
\subsection{Chip fabrication and characterisation}
The co-planar waveguide (CPW), resistance temperature detector (RTD) and heaters are deposited via photolithography on a circular glass substrate with a $25\,\mathrm{mm}$ diameter and $170\, (5)\,\mathrm{\mu m}$ thickness. The glass substrate is cleaned and a layer of S1813 photoresist is spincoated onto the surface. Sections of the photoresist are selectively exposed to UV light using a photolithography mask. The resist is then hardened and developed using Chlorobenzene and MF-319 developer. A thermal evaporator is used to deposit $5\,\mathrm{nm}$ of titanium, $200\,\mathrm{nm}$ of gold and a further $5\,\mathrm{nm}$ of titanium. Finally, the remaining photoresist is removed during lift off. 

\subsubsection{Insulation Layer}
\label{insulationlayer}
For PDMS coating, SYLGARD™ 184 Silicone Elastomer (PDMS) is mixed in a 7 to 1 ratio by weight and degassed for at least 30 minutes. A $5\,\mathrm{\mu m}$ layer of PDMS is spin coated onto the surface of the chip in two stages: $60\,\mathrm{s}$ at $300\,\mathrm{rpm}$ and $100\,\mathrm{rpm}\,\mathrm{s}^{-1}$ followed by $5\,\mathrm{min}$ at $6000\,\mathrm{rpm}$ and $500\,\mathrm{rpm}\,\mathrm{s}^{-1}$. 
For Parylene C (PaC) coating, a 2 $\mu$m layer is deposited with a Model 2010E Parylene deposition system.\\
The edges of the chips are protected during spin coating and chemical vapour deposition to avoid covering the contact pads.\\

\subsubsection{Chip Assembly}
The assembly of Q-BiC is carried out using Multicore Loctite RM89 solder paste which is heated to $180\,\mathrm{^{\circ}C}$. The alignment of the glass chip relative to the PCB is facilitated by the aluminium mount. 

\subsubsection{Autofluorescence Characterisation}
The auto-fluorescence of the different insulation layers was acquired by iteratively scanning the focal plane every 250 nm in z and averaging the photon counts of the ND free regions.

\subsubsection{PDMS Channels}
PDMS is mixed in a 7 to 1 ratio by weight and degassed for at least 30 minutes. The degassed mixture is poured into a mould and cured overnight in a $60\,\mathrm{^{\circ}C}$ oven. The PDMS wells are cut into shape and bonded to the assembled chips using a Plasma surface treatment machine.

\subsection{Finite element method simulation}
The RF delivery is simulated using COMSOL multiphysics RF module. The simulated geometry includes only the metal layer. The metallic layer is assumed to be infinitely thin with finite areal DC electrical resistance given by the measured thickness of gold. The frequency-domain simulation gives B field vector at every point in space, along with its relative phase. The magnetic field is elliptically polarised in general and the maximum amplitude is plotted in Fig. 2 (a) as the black dashed curve to indicate best possible RF drive with a well-aligned NV. The Rabi rate of the scanning NV probe (fixed orientation) at different spatial locations is found by coherently summing the projections of the magnetic field vector components into the plane orthogonal to the NV axis and computing the magnitude of the resulting complex number (blue curves in Fig. 2 (a)). The NV orientation is chosen such that the $B_1$ field is linearly polarised.

The heat generation simulation is performed with COMSOL multiphysics heat transfer module and AC/DC module. The on-chip heating simulation includes convection of the water in the PDMS well in the laminar regime, and thermal conduction through the objective and immersion oil. The back aperture of the objective is assumed to be at the incubator temperature and all other surfaces of Q-BiC are assumed to lose to heat via convective heat transfer through air at incubator temperature. The gold metal layer is assumed to be infinitely thin with area resistance given by the measured thickness of gold. The thin PDMS coating is found insignificant to heat conduction.

\subsection{Cell Culture}
In order to adhere and grow cells in the PDMS channels, the incubation channel was first sterilised by washing three times with 70\% ethanol (30\% water). Following three washes with phosphate buffered saline (PBS), the chip is incubated for a minimum of 1 hour at 37°C with the PDMS well filled with GelTrex solution (ThermoFisher, UK), which produces a basement membrane matrix on the inner surface of the well. Confluent HeLa cells are seeded at a concentration of $1\times10^{5}$ cells/mL for sufficient coverage after 12 hours. For cell viability and quantum sensing experiments, Leibovitz's L15 media supplemented with 10\% fetal bovine serum (FBS) (Sigma Aldrich, UK) was used.

Vesicle trajectories were captured on a Leica SP5 scanning microscope in widefield mode at a frame rate of 0.68 frames per second, and identified using the ImageJ plugin TrackMate \cite{ershov2022trackmate} (Supplementary Figure S6). The mean squared displacement, $\mathrm{MSD}(\tau)$, for each trajectory was calculated using,
$\mathrm{MSD}(\tau) = \langle \lvert \mathbf{{r}}(t+\tau)-\mathbf{{r}}(t)\rvert ^2\rangle$, where $\mathbf{r}$ is the projection of the vesicle position in the imaging plane and $\tau$ is the time interval.\\

\subsection{Nematode culture and strains}\label{WormCulture}
\textit{C. elegans} was grown on NGM agar plates seeded with \textit{E. coli} HB101, following standard procedures and maintained at 20 °C. \cite{Stiernagle2006}\\ \\
Wild-type strain was Bristol N2 \cite{Brenner1974}.\\ \\
Strain SX2635 \textit{mjIs228}: This strain carries a GFP-tagged immune-response-activated infection reporter (\textit{lys-3p::gfp}) in a N2 background. The question mark indicates that the position of the allele in the genome is currently unknown. Source: Miska lab (Jérémie Le Pen) \cite{LePen2018}.

\subsection{\textit{C. elegans} microinjection}
Young adult hermaphrodites were picked for microinjection following the standard procedure \cite{Mello1991}. Glass needles of inner diameter 0.5 $\mu$m (0.7 $\mu$m outer diameter, Eppendorf Femtotip ii) were filled with a 0.5 mg/mL nanodiamond suspension (100 nm, Hydroxy-terminated, FND Biotech) with microcapillary filling pipette tips (Microloader, Eppendorf).  ND suspensions were  sonicated in a water bath for 30 min at room temperature prior to loading. Worms were mounted on agar pads to immobilise them for injections and covered in mineral oil to slow desiccation. The specimen was then placed on an inverted microscope (Olympus IX71) equipped with a micromanipulator (InjectMan 4 Eppedorf) and microinjector (FemtoJet, Eppendorf). The injection pressure was set to 1900 hPa and between 3-4 pumps were injected into the distal arm of one or both gonads.

\subsection{Live immobilisation for quantum sensing} \label{liveimmob}
A custom aluminium mould with circular indentation containing one central triangular prism trench was filled with 10 w/v\% agarose (Sigma-Aldrich). After solidification, the agarose pads were taken out of the mould. 1.5 $\mu$L of 0.1 $\mu$m polystyrene latex beads (Polybead, Polysciences, Inc.) were pipetted into the triangular prism trench. A young adult hermaphrodite was washed in dH\textsubscript{2}O and added to the trench for nanoparticle-mediated immobilisation. The agarose pad was then inverted onto a glass slide for imaging or onto Q-BiC for quantum sensing. A custom acrylic ring was placed around the agarose pads to prevent spillage. The immobilised specimens were covered with custom lids and refilled with dH\textsubscript{2}O through an access port in the lid to avoid drying out.

\subsection{Image acquisition}
A Leica DM6 B fluorescent microscope was used to image live immobilised nematode specimens (10x Air Objective (NA = 0.32), 16 bit, 2048 (X) x 2048 (Y) pixels, exposure: 41.8 ms).\\
A Leica SP8 White Light inverted confocal microscope was used to image the stress reporter.\\

\subsection{\textit{C. elegans} stress assays}
\subsubsection{Population synchronisation}
Strain SX2635 was maintained according to standard protocols (\ref{WormCulture}).
Nematodes of the same age are required for a reliable readout of the \textit{lys-3} stress marker used, without bleaching or starvation to prevent stress. Synchronised populations of nematodes were obtained by gently picking around 15 young egg-laying adults (50-75 hours after hatching) with a platinum wire onto a fresh, seeded NGM plate. The young adults were kept at 20 °C for 1.5 hours to lay the eggs and then removed from the plate. The $\sim 50-100$ eggs were left to develop for 63 hours prior to exposure to stress.

\subsubsection{Stress exposure}
\subparagraph{Microwave stress} For microwave exposure stress assays, synchronised SX2635 nematodes were immobilised on Parylene C coated Q-BiCs for 20 min at different microwave powers (16.7 dBm, 17.4 dBm, 21.5 dBm, 22.3 dBm). Worms were recovered by adding dH\textsubscript{2}O, carefully lifting the agar pads and picking up the specimen with an eyelash pick onto a fresh plate.
\subparagraph{Temperature stress} For the temperature stress assay, synchronised SX2635 nematodes were placed on 50 mm seeded plates and inverted in an air-circulated incubator (Binder, Model KB 115) at 37 $^\circ$C for 90 min.
\subsubsection{Immobilisation for imaging}
Ony in cases when specimens were mounted for imaging the stress reporter, which does not require live animals, sodium azide was used.  4 w/v\% agarose pads (Sigma-Aldrich) were made between two microscope slides. After solidification, a drop of 1.5 $\mu$L sodium azide (5 \%, (0.75 M)) was added per worm on top of the agarose pad for anaesthetization. A nematode was transferred to the drop with a platinum wire, covered with another 1.5 $\mu$L of sodium azide (5 \%, (0.75 M)), then with a coverglass.
\subsubsection{Image acquisition}
For imaging of the strain SX2635, a 10x Air objective (NA = 0.4) was used (Base Zoom = 0.75, Gain = 600, 100 Hz, 16 bit, 512 (X) x 512 (Y) pixels). The laser power was set to 70 \% at the excitation wavelength of 488 nm. The collection ranges were 500-580 nm and 620-700 nm, respectively. Z-stacks were obtained from the imaging plane just below the pharynx to the imaging plane just above the pharynx with a Z-step size of 2.5 $\mu$m.
\subsubsection{Image analysis}
Fluorescent images of the stress reporter were analysed in Python. A thresholding filter was applied to remove background pixels and thus removing the need to normalise data with respect to the Z-stack height. Following, all pixel values of all frames in respective Z-stacks were summed up. To account for different worm volumes, the pixel sums were divided the respective worm volume. The worm volume was approximated from the length and cross-sectional surface area of the worm obtained from a single frame and assumes that a worm body can be approximated as a radially symmetric ellipsoid, where $l$ is the length of the worm, and $a$ half its width, and $A$ is the cross sectional area.
\begin{eqnarray*}
    V_{\mathrm{ellipsoid}} &=& \frac{4\pi}{3}a^2\frac{l}{2}\\
    A &=&  \pi a\frac{l}{2}\\
    \therefore \frac{2A^2}{\pi l} &=& \pi a^2\frac{l}{2}\\
    \therefore \frac{8}{3\pi}\frac{A^2}{l}&=& \frac{4}{3}\pi a^2\frac{l}{2}=V_{\mathrm{ellipsoid}}
\end{eqnarray*}

The worm length $l$ is computed using a custom fitting algorithm that detects the outline of the worm and computes the longest distance along the center line. This algorithm works by first taking the discrete cosine transform of the image, retaining only the 10 lowest frequency modes, and reconstructing a background image by taking the inverse transform.  This background image is subtracted from the original, and the resulting difference image is thresholded to segment the worm.  The resulting binary image is then skeletonised using the \textit{bwskel} function in MATLAB, which uses the medial axis transform method.  The end points of the skeleton are found morphologically using the \textit{bwmorph} function (Supplementary Figure S12). A custom function then computes the geodesic distance along the skeleton between all sets of endpoints and chooses the longest.
Worm volumes were calculated automatically but corrected by hand in cases when imaging artefacts (such as air inclusions near the worm body) interfered with reliable segmentation.
To account for variations in GFP expression due to external factors, each specimen was normalised with respect to the mean value of the control set of the day by subtraction, since the additional GFP signal in stressed specimens was not found to be specific to the regions in which GFP was expressed in control specimens (Supplementary Figure 13). Therefore, the values plotted in Fig. 4 (d) represent the relative stress of the specimen.

\bibliography{References}

\onecolumngrid
\pagebreak
\centering
\section*{Supplementary Materials for Q-BiC: A biocompatible integrated chip for \textit{in vitro} and \textit{in vivo} spin-based quantum sensing}

\setcounter{figure}{0}
\begin{figure*}[!htb]
\includegraphics[width=0.7\textwidth]{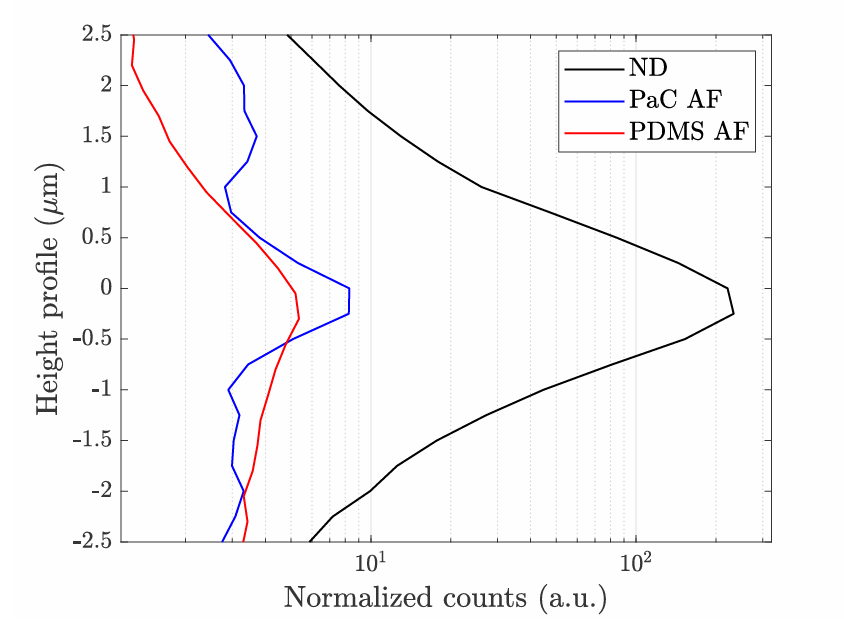}
\caption{Height profile of the insulation layer auto fluorescence in logarithmic scale. Height is centred around the ND focal plane. ND fluorescence and PaC data where taken on the same chip, PDMS was taken on a separate chip. Counts are normalised with respect to background counts to account for variations in the room light level and alignment conditions. Asymmetry in the data stems from the auto fluorescence of the glass.)
\label{Fig:Autofluorescence}}
\end{figure*}

\begin{figure*}[!htb]
\centering
\includegraphics[width=0.35\textwidth]{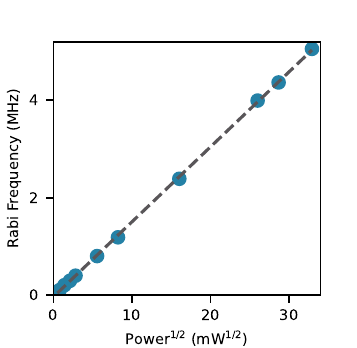}
\caption{Linear relationship between the Rabi frequency and the square root of the microwave power.  
\label{Fig:RabivsPower}}
\end{figure*}

\begin{figure*}[!htb]
\centering
\includegraphics[width=0.4\textwidth]{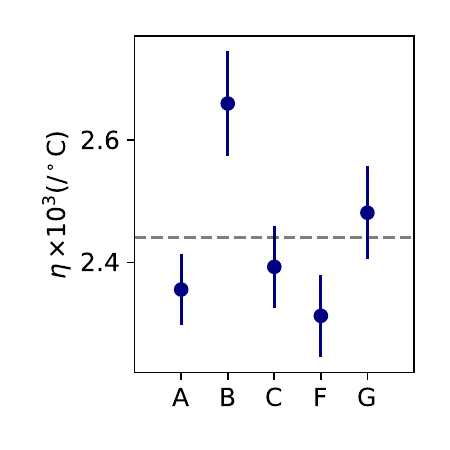}
\caption{Variations in $\eta$, the experimentally determined proportionality constant in the calibration equation $R(T)/R(T_{\mathrm{ref}}) = \eta (T-T_{\mathrm{ref}}) + 1$. This was characterised for five chips by varying the temperature of an incubator box. The temperature of the incubator was verified using commercially available thermocouples and RTDs. There was seen to be less than 5\% variation between chips.
\label{Fig:RTD calibration}}
\end{figure*}

\newpage

\begin{figure*}[!htb]
\centering
\includegraphics[width=0.7\textwidth]{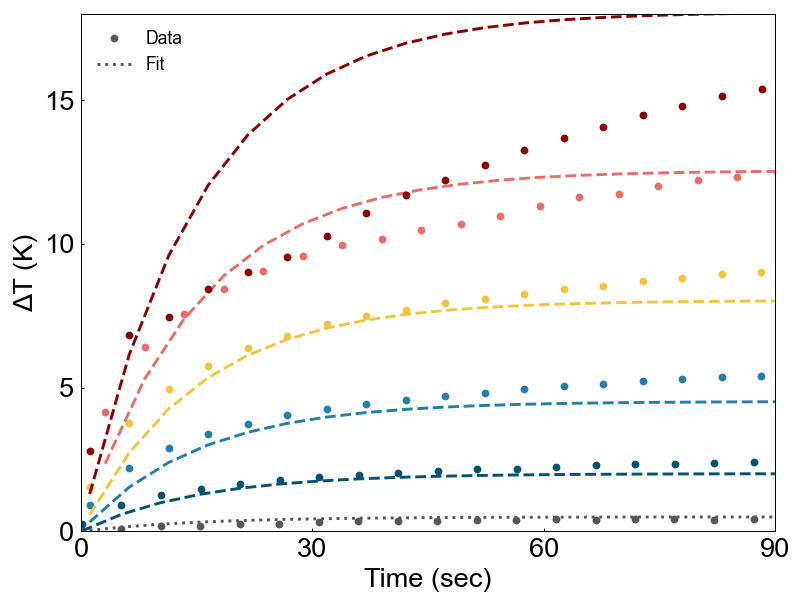}
\caption{
Fitting the heater temperature ramps (dots) using equation 3 (dashed curves) using only $C_v$ and \textit{k} as fitting parameters shows good agreement with $C_v = 0.26 \pm 0.01 \mathrm{J/K}$ and $k = 0.0175 \pm 0.0002 \mathrm{J/K/s}$. This model only takes into account heat loss due to thermal conduction which may account for the discrepancy at higher voltages.
\label{Fig:heating theory}}
\end{figure*}

\begin{figure*}[!htb]
\centering
\includegraphics[width=0.7\textwidth]{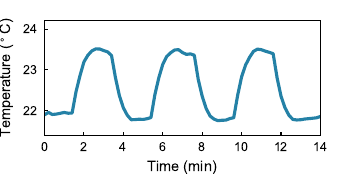}
\caption{ Demonstration of Q-BiCs ability to change temperature over short timescales. The temperature was alternated every 2 minutes with an difference of $1.5\,^\circ \mathrm{C}$.
\label{Fig:Quicktempstepping}}
\end{figure*}

\begin{figure*}[!htb]
\includegraphics[width=0.7\textwidth]{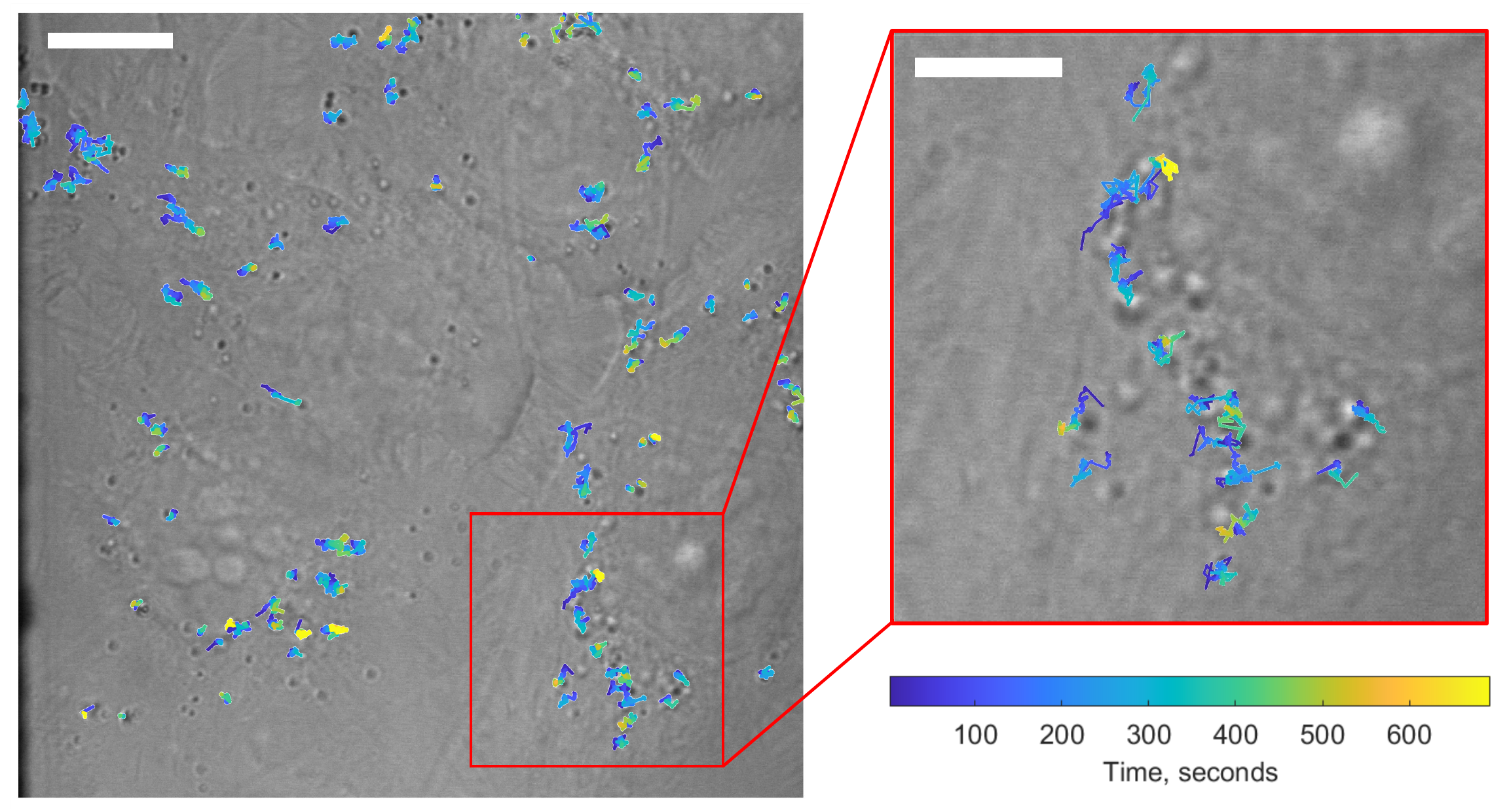}
\caption{Example trajectories of intracellular vesicles overlayed on the first frame of the time lapse videos, scalebar = 10 $\mu$m (with inset showing motions within a single cell, scalebar = 5 $\mu$m). Colour represents length of the trajectory over time. 
\label{Fig:CellViability}}
\end{figure*}

\begin{figure*}[!htb]
\includegraphics[width=0.5\textwidth]{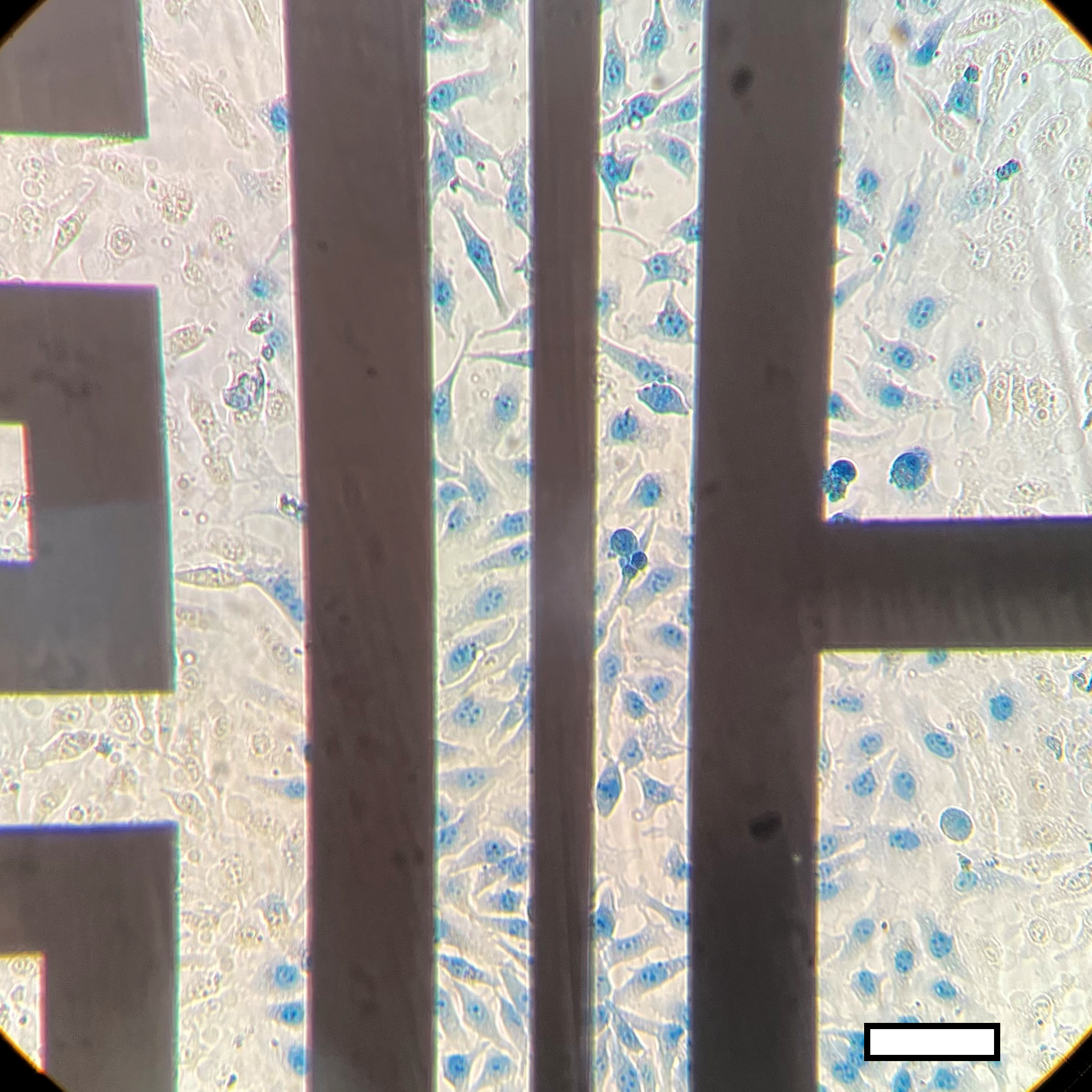}
\caption{Trypan blue stain retention in cells (indicative of cell death) in close proximity ($<$ 300 $\mu$m) to the microwave antenna following 25.6 dBm power input. Scalebar = 100 $\mu$m
\label{Fig:DeadCells}}
\end{figure*}

\begin{figure*}[!htb]
\centering
\includegraphics[width=0.6\textwidth]{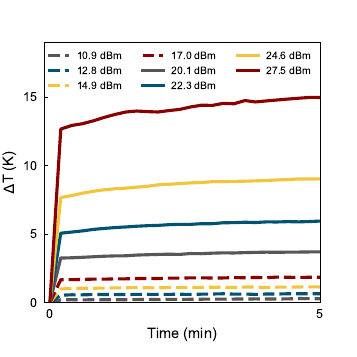}
\caption{Temperature increase measured by on-chip RTD for different microwave powers. The temperature is seen to increase significantly in the first 12 seconds and remains stable after this point.
\label{Fig:MWheating}}
\end{figure*}

\begin{figure*}[!htb]
\includegraphics[width=1\textwidth]{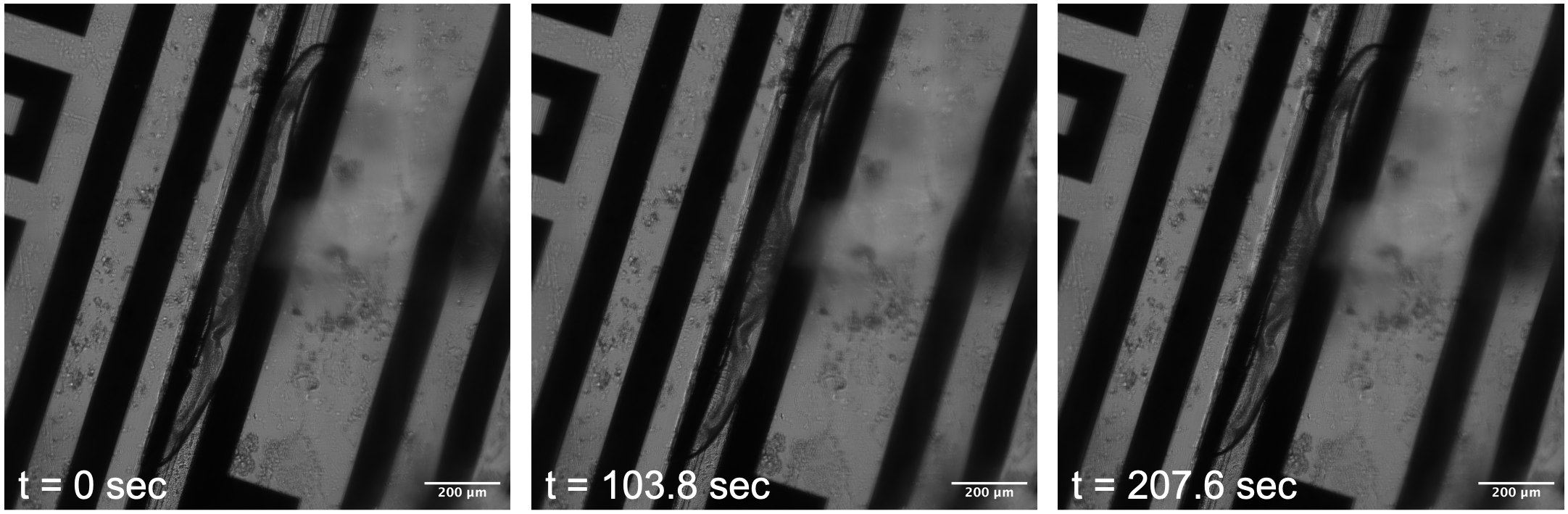}
\caption{Images of an immobilised worm at three different time points to show successful live immobilisation along the microwave line of a Q-BiC. Video of immobilised, non-anaesthetised worm shown as separate supplementary. 
\label{Fig:WormImmob}}
\end{figure*}

\begin{figure*}[!htb]
\includegraphics[width=1\textwidth]{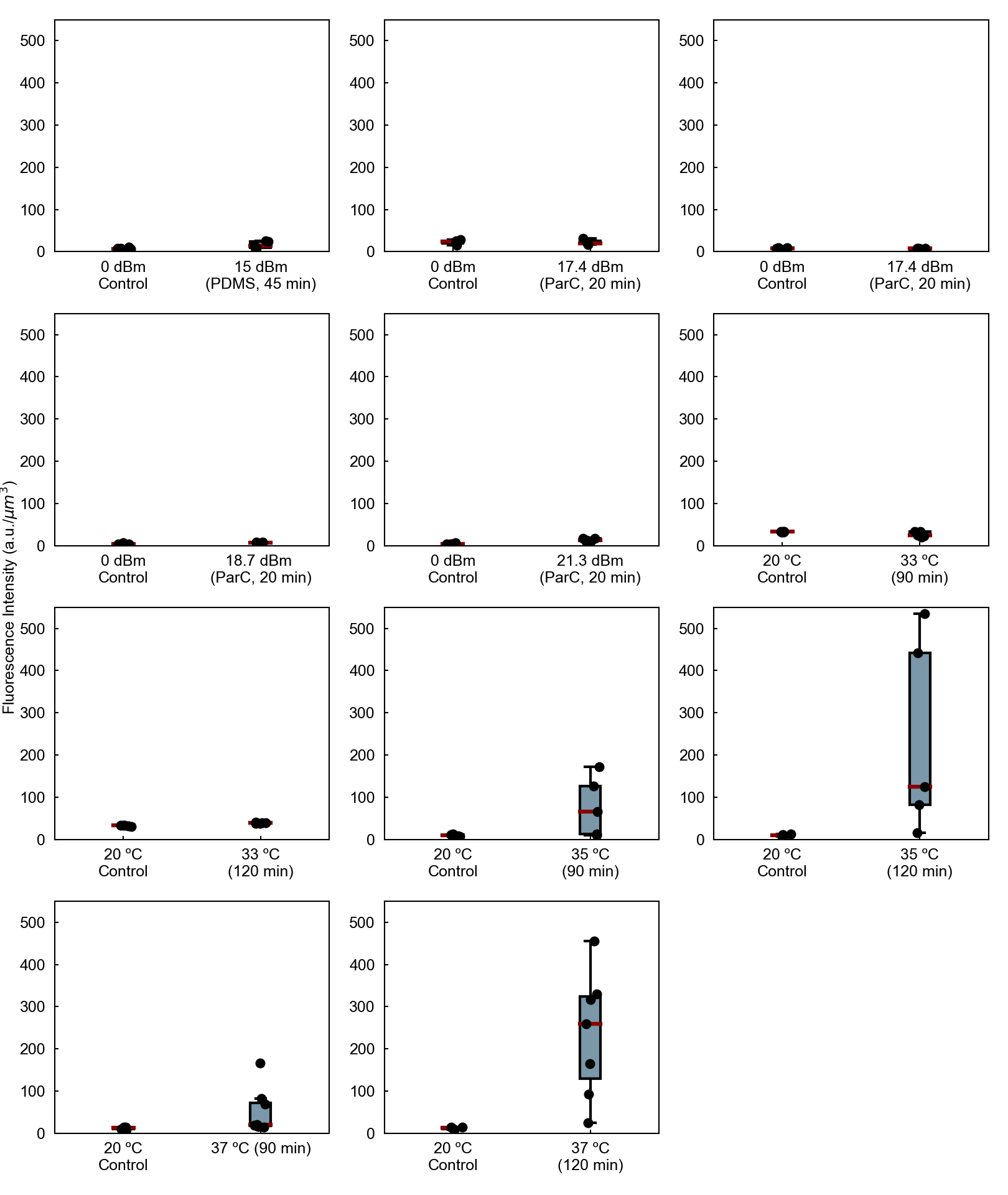}
\caption{Fluorescent intensity of GFP expressed in stressed SX2635 \textit{C. elegans} normalised with respect to their individual volumes. Each subplot shows a stress condition together with its respective control group. (PDMS = PDMS coated Q-BiC were used, ParC = Parylene C coated Q-BiCs were used.) 
\label{Fig:WormStressSupp}}
\end{figure*}

\begin{figure*}[!htb]
\includegraphics[width=1 \textwidth]{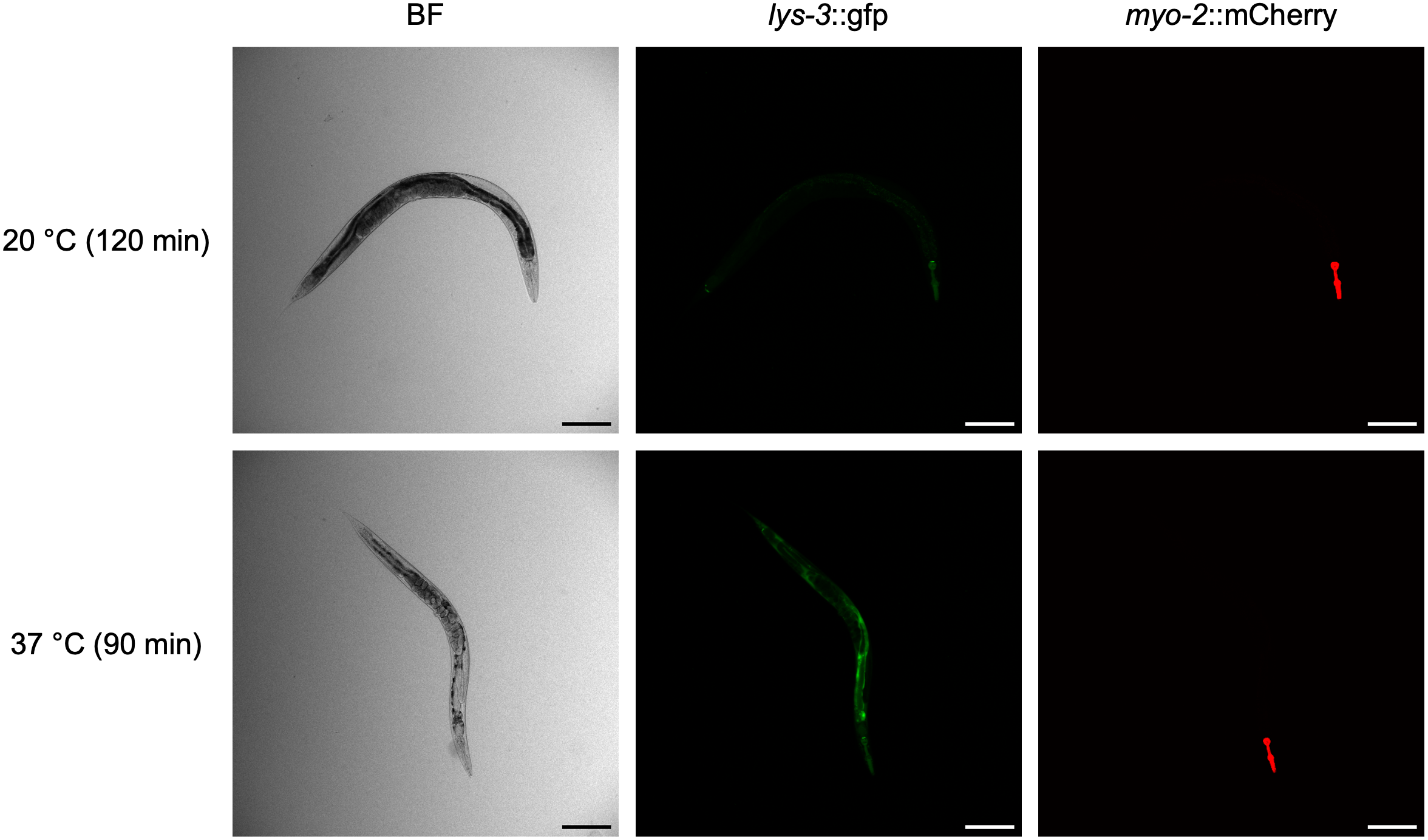}
\caption{Confocal images of a \textit{C. elegans} sample control (top row, 20 ºC, 120 min) and stress exposure (bottom row, 37 ºC, 90 min). Shown are three colour channels: brightfield (BF) (single z-slice), green fluorescent protein (GFP) (sum z-projection), mCherry (from left to right)(single z-slice). Scalebar = 200 $\mu$m. 
\label{Fig:SX2635}}
\end{figure*}

\begin{figure*}[!htb]
\includegraphics[width=0.7\textwidth]{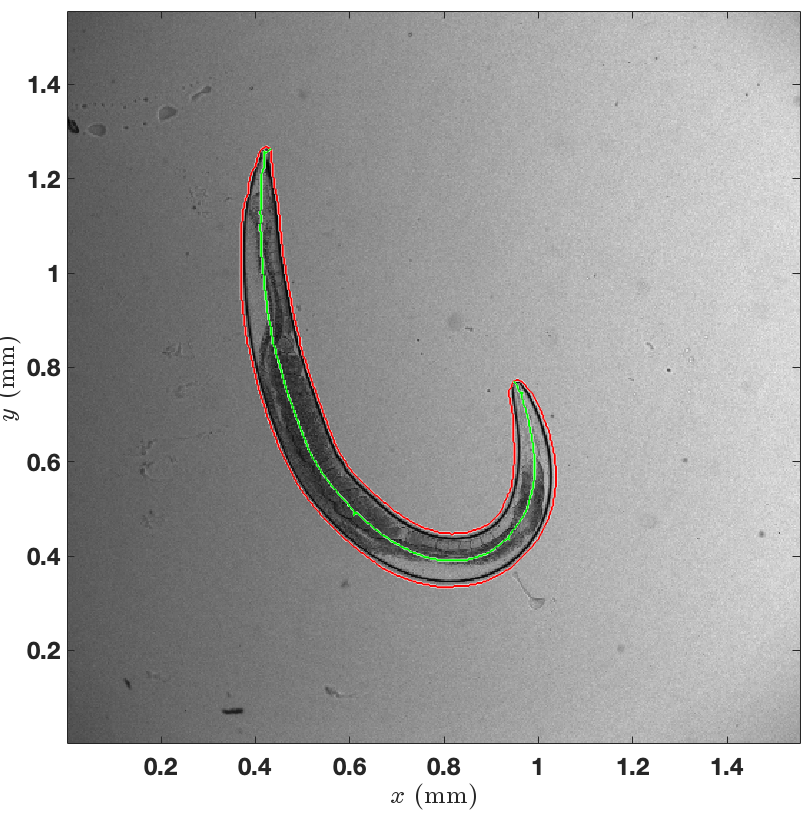}
\caption{Confocal image of a \textit{C. elegans} sample with an outline fit (red) to obtain the surface area. The fit to obtain the worm length is shown in green. Shorter green paths are discarded and only the single longest path is used for calculation of the worm volume. 
\label{Fig:WormFit}}
\end{figure*}

\begin{figure*}[!htb]
\includegraphics[width=1 \textwidth]{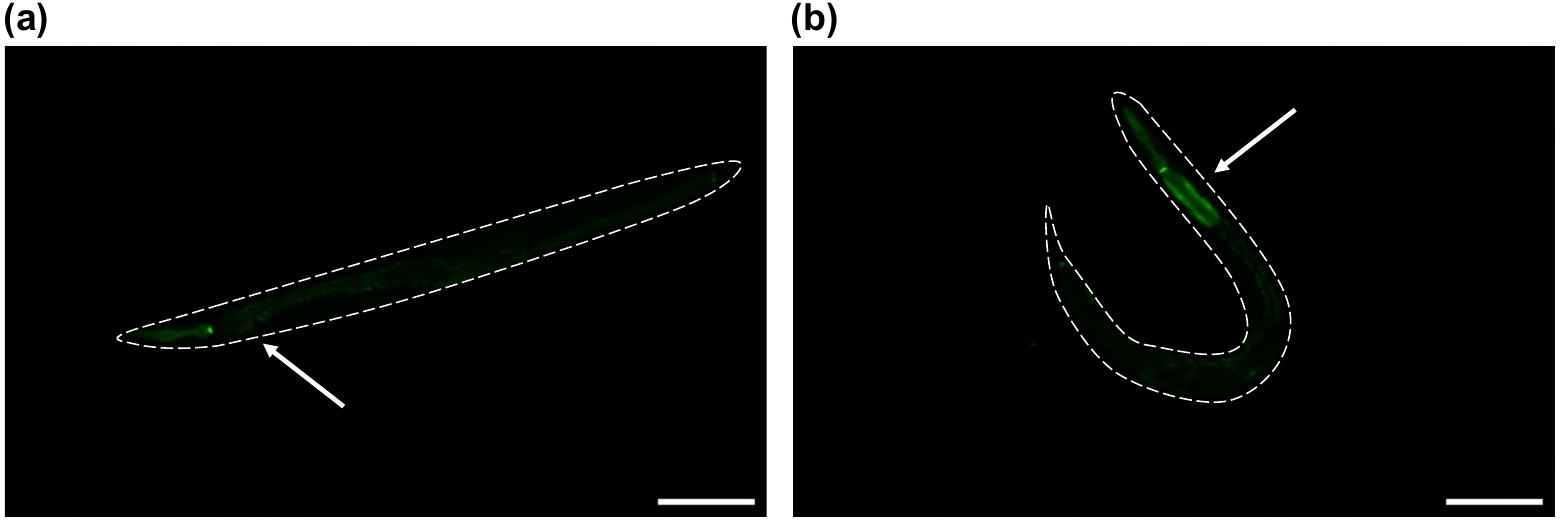}
\caption{Confocal images of green fluorescent protein (GFP) (sum z-projection) of \textit{C. elegans} samples with the worm body outline shown as a dashed white line. \textbf{(a)} shows the control and \textbf{(b)} shows the a sample stressed at 25.4 dBm microwave exposure for 45 min. The white arrow highlights the same region in the two samples. It can be seen that the additional GFP signal in the stressed sample is not found to be specific to regions with expressed GFP in the control. Scalebar = 200 $\mu$m}
\end{figure*}

\FloatBarrier
\begin{center}
\begin{table*}[h]
\caption{\textit{C. elegans} lethality measured after overnight recovery post stress exposure.}
\begin{tabular}{c||c|c|c|c|c} 
Stress Type & n (Alive+Dead) & Alive & Dead & Miscellaneous & Lethality \\
\hline
33 °C, 90 min & 8 & 8 & 0 & 0 & 0 \\
33 °C, 120 min & 8 & 8 & 0 & 0 & 0 \\
Control & 9 & 9 & 0 & 0 & 0 \\
\hline
35 °C, 90 min & 14 & 13 & 1 & 6 & 0.071 \\
35 °C, 120 min & 19 & 18 & 1 & 1 & 0.053 \\
Control & 15 & 15 & 0 & 2 & 0 \\
\hline
15 dBm, 45 min (PDMS) & 8 & 6 & 2 & 0 & 0.25 \\
Control & 10 & 10 & 0 & 0 & 0 \\
\hline
16.7 dBm, 20 min (Parylene C)& 4 & 4 & 0 & 0 & 0 \\
Control & 6 & 6 & 0 & 1 & 0 \\
\hline
16.7 dBm, 20 min (Parylene C)& 7 & 6 & 1 & 0 & 0.14 \\
Control & 7 & 7 & 0 & 0 & 0 \\
\hline
17.4 dBm, 20 min (Parylene C)& 5 & 5 & 0 & 0 & 0 \\
21.5 dBm, 20 min (Parylene C)& 5 & 5 & 0 & 0 & 0 \\
22.3 dBm, 20 min (Parylene C)& 6 & 0 & 6 & 0 & 1 \\
Control & 6 & 6 & 0 & 2 & 0 \\
\end{tabular}
\end{table*}
\end{center}
\FloatBarrier

\begin{figure*}[!htb]
\includegraphics[width=0.8\textwidth]{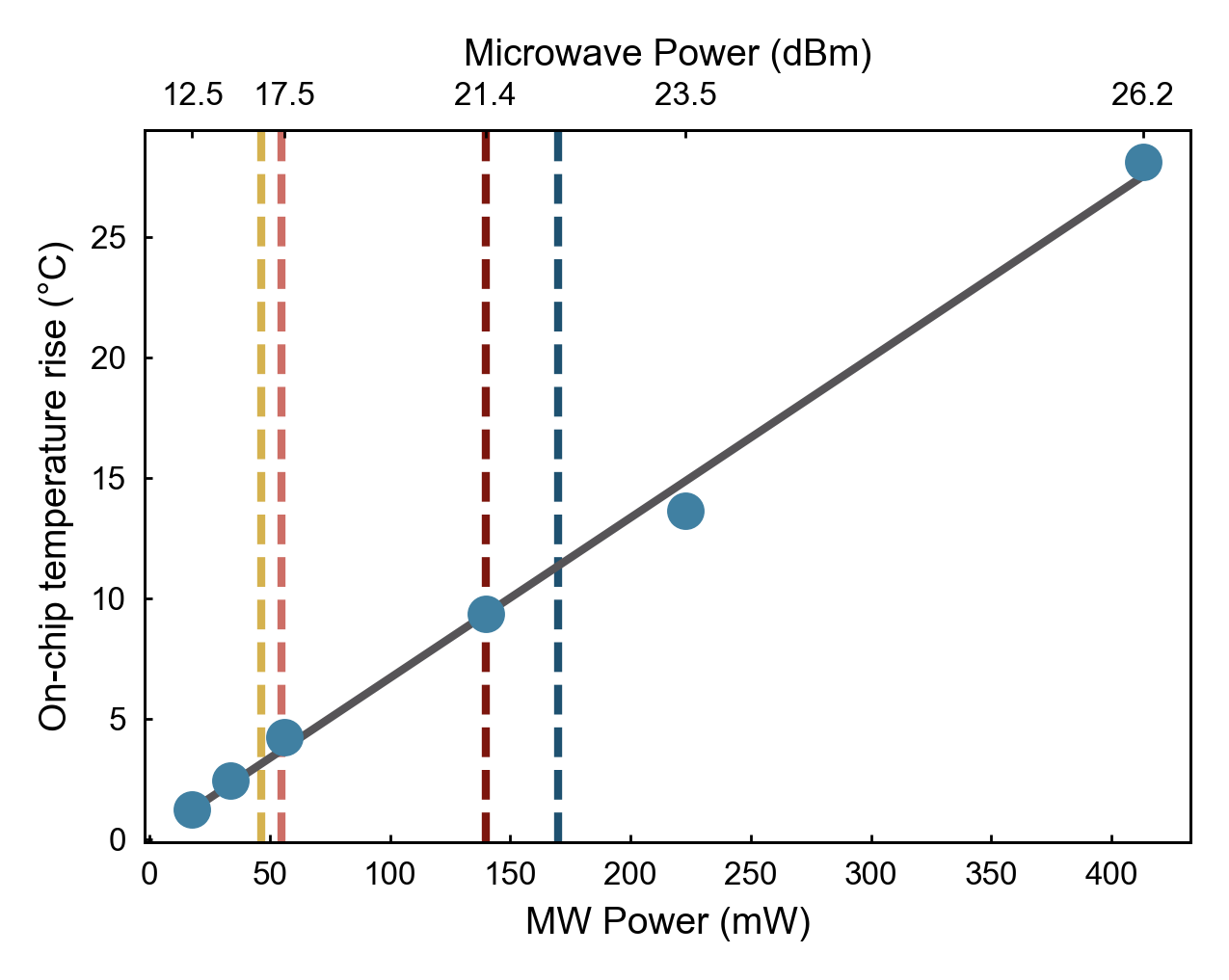}
\caption{The on-chip temperature rise of a Parylene C coated Q-BiC observed for different microwave powers (solid blue circles) with the best linear fit (grey curve). The microwave powers used for stress sensing in \textit{C. elegans} (16.7 dBm, 17.4 dBm, 21.5 dBm and 22.3 dBm) are highlighted as dashed lines.  
\label{Fig:SuppMWPowerStepping}}
\end{figure*}

\end{document}